\documentclass[twocolumn]{aastex63}
\usepackage{amssymb, amsfonts, amsthm}

\usepackage{amsmath}
\usepackage{floatrow}
\usepackage{appendix}
\usepackage{chngcntr}
\newcommand{\hi}{H{\sc i}}

\submitjournal{APJL}

\shorttitle{post-EoR upper limits}
\shortauthors{Chakraborty et al.}
\graphicspath{{./}{figures/}}
\begin{document}

\title{First multi-redshift limits on post-Epoch of Reionization (post-EoR) 21 cm signal from ${\bf z = 1.96 - 3.58}$ using uGMRT}

\correspondingauthor{Arnab Chakraborty}
\email{phd1601121009@iiti.ac.in, arnab.phy.personal@gmail.com}

\author[0000-0002-0786-7307]{Arnab Chakraborty}
\affiliation{Discipline of Astronomy, Astrophysics and Space Engineering, Indian Institute of Technology Indore, Indore 453552, India}

\author{Abhirup Datta}
\affiliation{Discipline of Astronomy, Astrophysics and Space Engineering, Indian Institute of Technology Indore, Indore 453552, India}

\author{Nirupam Roy}
\affiliation{Department of Physics , Indian Institute of Science,Bangalore 560012,India}

\author{Somnath Bharadwaj}
\affiliation{Department of Physics $\&$ Centre for Theoretical Studies,IIT Kharagpur,Kharagpur 721302,India}

\author{Tirthankar Roy Choudhury}
\affiliation{National Centre For Radio Astrophysics,Post bag 3,Ganeshkhind,Pune 411007,India}

\author{Kanan K. Datta}
\affiliation{Department of Physics, Presidency University, 86/1 College Street, Kolkata-700073, India}

\author{Srijita Pal}
\affiliation{Department of Physics $\&$ Centre for Theoretical Studies,IIT Kharagpur,Kharagpur 721302,India}

\author{Madhurima Choudhury}
\affiliation{Discipline of Astronomy, Astrophysics and Space Engineering, Indian Institute of Technology Indore, Indore 453552, India}

\author{Samir Choudhuri}
\affiliation{School of Physics and Astronomy, Queen Mary University of London, London E1 4NS, UK}

\author{Prasun Dutta}
\affiliation{Department of Physics, IIT (BHU) Varanasi, 221005 India}

\author{Debanjan Sarkar}
\affiliation{Department of Physics, Ben-Gurion University of the Negev, Be’er Sheva - 84105, Israelel}

\begin{abstract}
 Measurement of fluctuations in diffuse \hi\ 21 cm background radiation from the post-reionization epoch ($z \leq 6$)  is a promising avenue to  probe the large scale structure of the Universe and understand the evolution of galaxies. We observe the European Large-Area ISO Survey-North 1 (ELAIS-N1) field at 300-500 MHz using the upgraded Giant Meterwave Radio Telescope (uGMRT) and employ the ‘foreground avoidance’ technique  to estimate the \hi\ 21 cm power spectrum in the redshift range $z = 1.96-3.58$.  Given the possible systematics that may remain in the data, we find the most stringent upper limits on the spherically averaged 21 cm  power spectra at $k \sim 1.0$ $\mathrm{Mpc}^{-1}$ are  (58.87 $\mathrm{mK})^{2}$, (61.49 $\mathrm{mK})^{2}$, (60.89 $\mathrm{mK})^{2}$, (105.85 $\mathrm{mK})^{2}$  at $z = 1.96,2.19,2.62$ and $3.58$, respectively. We use this to constrain the product of neutral \hi\ mass density ($\Omega_{\mathrm{H{\sc I}}}$) and \hi\ bias ($b_{\mathrm{H{\sc I}}}$) to the underlying dark matter density field, [$\Omega_{\mathrm{H{\sc I}}} b_{\mathrm{H{\sc I}}}$], as 0.09,0.11,0.12,0.24 at $z=1.96,2.19,2.62,3.58$, respectively. To the best of our knowledge these are the first limits on the \hi\ 21 cm power spectra  at the redshift range $z = 1.96 - 3.58$ and would play a significant role to constrain  the models of galaxy formation and evolution.

\end{abstract}


\section{Introduction} \label{sec:intro}
The redshifted 21 cm line emission from neutral hydrogen (H{\sc i}) provides a rich tool to map the Universe in 3D. The majority of \hi\ is ionized by ultra violet radiation emanating from early galaxies during a period $z\sim 15-6$ (Epoch of Reionization; EoR) \citep{Madau1997ApJ...475..429M}. However, below $z\sim6$ (post-EoR epoch) H{\sc i} is contained within dense clumps, which are self-shielded from the ionizing radiation. These dense clumps are strongly correlated with over density of matter where abundant of H{\sc i} are being self-shielded.  The distribution of H{\sc i} is intimately connected to matter distribution of the Universe and hence help to understand the large-scale structures at intermediate redshifts in the post-EoR era ($z\lesssim 6$) \citep{Bull2015AApJ...803...21B}. Along with that measurement of  post-EoR \hi\ 21 cm power spectrum can be used to study Baryon Acoustic Oscillations (BAO) and the equation of state of the dark energy \citep{Chang2008PhRvL.100i1303C,Somnath2009PhRvD..79h3538B,Bull2015AApJ...803...21B}.

There are several methods exist to measure the neutral H{\sc i} mass density ($\Omega_{\mathrm{H{\sc I}}}$) from $z$ = 0 to 6, starting from Lyman-alpha line absorption feature in a distant quasar spectra by intervening H{\sc i} region  at $z \gtrsim 1.5$ \citep{Prochaska2009ApJ...696.1543P}  to detecting 21 cm line emission from individual Galaxies at $z \lesssim 0.1$ \citep{Zwann2005MNRAS.359L..30Z,Martin2010ApJ...723.1359M}. It is extremely difficult to detect individual galaxies at $z \gtrsim 0.3$, which requires very deep integration time \citep{Nissim2016ApJ...818L..28K}. Nonetheless, one can obtain information about average properties of neutral $gas$ by co-adding the H{\sc i} 21 cm signal from large number of galaxies with known redshifts (`stacking') to boost the signal-to-noise ratio \citep{Lah2007MNRAS.376.1357L,Nissim2016ApJ...818L..28K,Apurba2019ApJ...882L...7B}. However, this method has so far been applied to low redshifts, $z \lesssim 0.4$. Another unique  technique is 3D `intensity mapping', which measure the fluctuations in the diffuse 21 cm background radiation \citep{Bharadwaj_Nath2001JApA...22...21B,Loeb2008PhRvL.100p1301L,Chang2010Natur.466..463C,Bull2015AApJ...803...21B}.  Previous studies use cross-correlation of the single-dish \hi\ 21 cm intensity map with deep galaxy survey and put constraint on [$\Omega_{\mathrm{H{\sc I}}} b_{\mathrm{H{\sc I}}}$] $\sim$ [$0.6^{+0.23}_{-0.15}] \times 10^{-3}$ at $z \simeq 0.8$ \citep{Masui2013ApJ...763L..20M,Switzer2013MNRAS.434L..46S}. \citet{AbhikB2011MNRAS.418.2584G} measure the fluctuations in the faint \hi\ 21 cm background using GMRT for the first time and put upper limit on [$ \bar{x}_{\mathrm{HI}} b_{\mathrm{H{\sc I}}}$] $\leq$ 2.9 ([$\Omega_{\mathrm{H{\sc I}}} b_{\mathrm{H{\sc I}}}$] $\leq$ 0.11) at $z \sim 1.32$, where $ \bar{x}_{\mathrm{HI}}$ is the mean neutral fraction.

The major challenge in detecting H{\sc i} power spectrum is the presence of the bright synchrotron radiation from galactic and extragalactic sources. Several novel techniques have been developed  to remove foregrounds \citep{AbhikA2011MNRAS.411.2426G,Liu2012MNRAS.419.3491L,Masui2013ApJ...763L..20M,Wolz2017MNRAS.470.3220W,Anderson2018MNRAS.476.3382A}.  Also foregrounds can be avoided in Fourier space ($\mathbf{k_{\perp}}, k_{\parallel}$), where the smooth foregrounds coupled with instrument response are localized in a `wedge' shape region \citep{Abhi2010ApJ...724..526D,Parsons2012AApJ...756..165P}. This method is being widely used to detect the H{\sc i} 21 cm signal from EoR \citep{Kolopanis2019ApJ...883..133K, Trott2020MNRAS.493.4711T}. We  follow the `foreground avoidance' method to estimate the H{\sc i} power spectrum at redshifts $z=1.96,2.19,2.62,3.58$ with uGMRT.   We also put the upper limits on the product of $\Omega_{\mathrm{H{\sc I}}}$ and  $b_{\mathrm{H{\sc I}}}$ at each redshift. The quantity, [$\Omega_{\mathrm{H{\sc I}}} b_{\mathrm{H{\sc I}}}$], contains information about the host dark matter haloes of H{\sc i} gas and  determines the amplitude of the expected H{\sc i} power spectrum. It is essential to put tight constraint on these parameters using observations  to predict the uncertainties in measuring H{\sc i} power spectrum for current and future telescops \citep{Somnath2005MNRAS.356.1519B,Battye2012arXiv1209.1041B,Hamsa2015MNRAS.447.3745P}.

\section{Observation and analysis} \label{sec:obs}

We observed  the ELAIS-N1 field ($\alpha_{2000}=16^{h}10^{m}1^{s} ,\delta_{2000}=54^{\circ}30'36\arcsec$ ) using uGMRT in GTAC cycle-32 during May-June 2017 at 300 - 500 MHz for a total time of 25 hours (including calibrators) over four nights.  ELAIS-N1 is a well known field in the northern sky  at high galactic latitude ($b=+44.48^{\circ}$) and previously studied at different frequencies (see \citealt{ArnabA2019MNRAS.487.4102C} and references therein). The data was taken  with  a time resolution of $2$ sec and frequency resolution of $24$ KHz using upgraded digital backend correlator \citep{Yashwant2017CSci..113..707G}. The detail analysis of editing bad data, calibration and imaging are mentioned in \citet{ArnabB2019MNRAS.490..243C}. However, here we did not average the data across frequency to get the maximum $k_{\parallel}$ modes, which is inversely proportional to the frequency resolution \citep{Morales2004ApJ...615....7M}.  We have used a mask during imaging, generated via P{\tiny Y}BDSF \footnote{\url{https://www.astron.nl/citt/pybdsf/}}, to ensure that imaging artifacts do not propagate into the model during direction-independent (DI) self-calibration. This results into building more accurate sky model consisting of bright compact sources and allows for the mitigation of calibration errors in self-calibration.  Also, during the self-calibration, we have excluded the shorter baselines ($<$1.5k$\lambda$), where the diffuse emission is most sensitive. Hence, there will not be any significant suppression of the diffuse emission and the 21 cm signal during self-calibration and subsequent sky-model subtraction \citep{Patil2016MNRAS.463.4317P}.  After getting the final image we found that there are 3728 components (compact sources) present in the model. 
Then  we have subtracted this point-source model from the calibrated visibility data, using  \textit{UVSUB} in {\tiny CASA}. This residual  data is being used for power spectrum analysis. We do not attempt to model and subtract the diffuse foreground emissions in this  analysis to avoid any suppression or loss of the diffuse 21 cm signal. 

%and done in  {\tiny CASA} \footnote{See: \url{https://casa.nrao.edu}; \citep{McMullin2007ASPC..376..127M} }

\section{Power spectrum estimation} \label{sec:PS_estimation}

  The  Fourier transformation of a visibility along the frequency direction to the $\eta$-domain  is given as \citep{Morales2004ApJ...615....7M} 

\begin{equation}
    V (\bold U,\eta) = \int V (\bold U,\nu)  S(\nu) W(\nu) e^{i2\pi \nu \eta} d\nu, 
    \label{eqn_2}
\end{equation}

where, $ V (\bold U,\eta)$ is the measured visibility of a baseline $\bold U$ as a function of frequency ($\nu$) and $W(\nu)$ is the Blackman-Harris (BH) window function used to control the visibility spectrum in the $\eta$-domain, $S(\nu)$ contains frequency dependent sample weights that result from  flagging of frequency channels due to radio frequency intereference (RFI).  We use one-dimensional `$\it{CLEAN}$' \citep{Hogbom1974A&AS...15..417H}  to deconvolve the kernel that results from the Fourier conjugate of the product of  [$W(\nu) S(\nu)$] and obtain the final spectra \citep{Parsons2009AJ....138..219P}.

The cylindrically averaged  power spectrum can be estimated with the use of proper scaling factor as \citep{Morales2004ApJ...615....7M,Parsons2012AApJ...756..165P} :
\begin{equation}
    P(\bold k_{\perp}, k_{\parallel}) = \Big(\frac{\lambda^{2}}{2k_{B}}\Big)^{2}    \Big(\frac{X^{2}Y}{\Omega B}\Big)  |V (\bold U,\eta)|^{2}, %\footnote[1]{We use cosmological parameters of the Planck 2018 analysis \citep{Planck2018arXiv180706209P}}
    \label{ps_eqn}
\end{equation}

with,
 \begin{align}
     \bold k_{\perp} = \frac{2\pi }{D(z)} \bold U, &&  k_{\parallel} = \frac{2\pi  \nu_{21} H_{0} E(z)}{c(1+z)^{2}} \eta, %\footnote[1]{We use cosmological parameters of the Planck 2018 analysis \citep{Planck2018arXiv180706209P}}
 \end{align}
where $\lambda$ is the wavelength corresponding to the band-center, $K_{B}$ is the Boltzmann constant, $\nu_{21}$ is the rest-frame frequency of the 21 cm spin flip transition of H{\sc i}, $z$ is the redshift to the observed frequency, $\Omega$ is sky-integral of the squared antenna
primary beam response, B is the bandwidth and $X^{2}Y$ is a redshift  dependent scalar to convert angle and frequency to cosmological length scales \citep{Morales2004ApJ...615....7M}. Here, $D(z)$ is the transverse comoving distance at redshit $z$, $H_{0}$ is the Hubble parameter and $E(z) \equiv [\Omega_{m}(1+\textit{z})^{3} +  \Omega_{\Lambda}]^{1/2}$.  $\Omega_{\mathrm{m}}$ and $\Omega_{\Lambda}$ are matter and dark energy densities, respectively.  In this work, we use the best fitted cosmological parameters of the Planck 2018 analysis \citep{Planck2018arXiv180706209P}. The power spectrum $P(\bold k_{\perp}, k_{\parallel})$ is in units of $\mathrm{K}^{2} \mathrm{Mpc}^{3}$.  %{\bf The intrinsic chromatic response of the instrument confines the foregrounds within a `wedge' shape region in this 2D ($\bold k_{\perp}, k_{\parallel}$) space and the H{\sc i} signal is searched for in a `cosmological signal window', devoid of foreground contamination above the `foreground wedge'.} %This `cosmological signal window' is identical to the `EoR window' at higher redshift ($z \gtrsim 6$) \citep{Abhi2010ApJ...724..526D, Parsons2012bApJ...756..165P,Vedantham2012ApJ...745..176V}.} 
We spherically averaged $P(\bold k_{\perp}, k_{\parallel})$ in $k$-bins  and estimate the dimensionless power spectrum as \citep{Abhi2010ApJ...724..526D}: 
\begin{equation}
   \Delta^{2} (k) = \frac{k^{3}}{2\pi^{2}} <P(\bold k)>,  
\end{equation}
where $k = \sqrt{k_{\perp}^{2} + k_{\parallel}^{2}}$.

However, correlating a visibility with itself, as in Eqn. \ref{ps_eqn},  will result in a positive bias due to the noise present in the data \citep{Somnath2003JApA...24...23B}. %We follow the same procedure and use the `$\it{CLEANPS}$' pipeline as described in Chakraborty,A+ 2020 to avoid the positive noise bias in power spectrum estimation.
 To avoid the positive noise bias, we cross correlate all the visibilites among each other within a $uv$-cell, whose dimension is inverse of the half-power beam width of the primary beam ($\theta_{\mathrm{HPBW}}$).  The off-diagonal terms of the correlation matrix for each $uv$-cell are expected to be free of noise bias  and the average of those terms is being quoted as the estimated power corresponding to that cell. The method of correlating visibilites  within a $uv$-cell to measure the post-EoR  21 cm power spectrum was first proposed by \citet{Bharadwaj_Sethi2001JApA...22..293B} and further discussed  in \citet{Somnath2003JApA...24...23B, Somnath2005MNRAS.356.1519B}.

 The power spectrum uncertainties are estimated by dividing the noise power by the square-root of the number of modes averaged together in estimating the power spectrum \citep{Tegmark1997PhRvD..55.5895T}.  We calculate the noise power by subtracting the average of the off-diagonal from the diagonal (self-correlation of visibilities) components for each  $uv$-cells. The estimated noise at the highest $k_{\parallel}$ modes (above the horizon limit) is expected to be dominated by thermal variance  \citep{Kolopanis2019ApJ...883..133K}.

\begin{figure}[htp]
    \centering
   
%\begin{tabular}{cc}
       \includegraphics[width=\columnwidth,height=3.in]{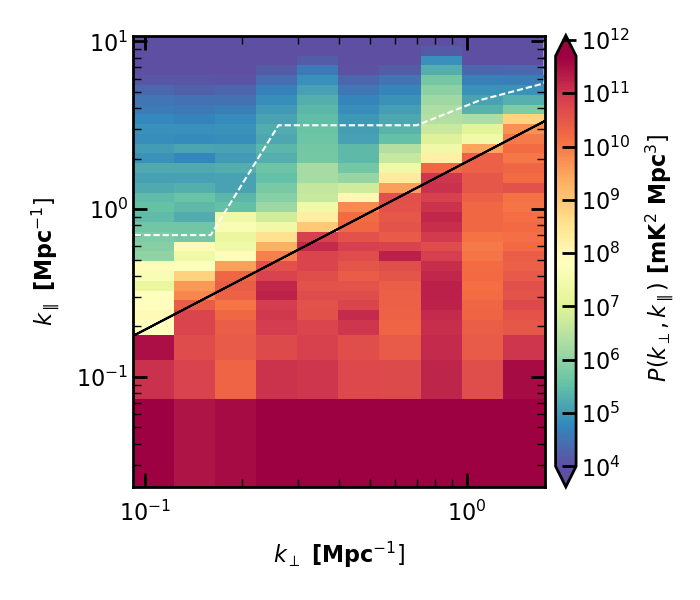} 
%        \includegraphics[width=\columnwidth,height=2.5in]{figures/3d_ps_with_HI.png} 
       
%\end{tabular}      
         
    \caption{The cylindrically averaged power spectrum at $z=3.58$. The black line is the horizon delay limit. The modes bounded by the white dashed lines are used to estimate spherically averaged power spectrum.}
    \label{PS_310}
\end{figure}

%This formalism is based on the idea that correlation of two visibilities in each $uv$-cell measure the same cosmological mode to the leading order (hence correlated) but have uncorrelated noise realization \citep{Somnath2003JApA...24...23B, Ali2008MNRAS.385.2166A,Abhik2012MNRAS.426.3295G,somnath2019MNRAS.483.5694B}.  

\section{Results} \label{sec:results}
The whole 200 MHz bandwidth data are divided into 8 MHz sub-bands and we only use projected baselines upto 2 km to estimate the \hi\ 21 cm power spectrum for each of these sub-bands. This choice allow us to restrict to minimal baseline migration over the sub-band. Also, the analysis using these sub-bands ensures the signal ergodicity within the volume, i.e, the cosmological signal does not evolve significantly over the bandwidth \citep{Kanan2014MNRAS.442.1491D,Trott2020MNRAS.493.4711T,Rajesh2020MNRAS.494.4043M}.

\begin{figure}[htp]
    \centering
   
%\begin{tabular}{cc}
        
       \includegraphics[width=\columnwidth,height=2.8in]{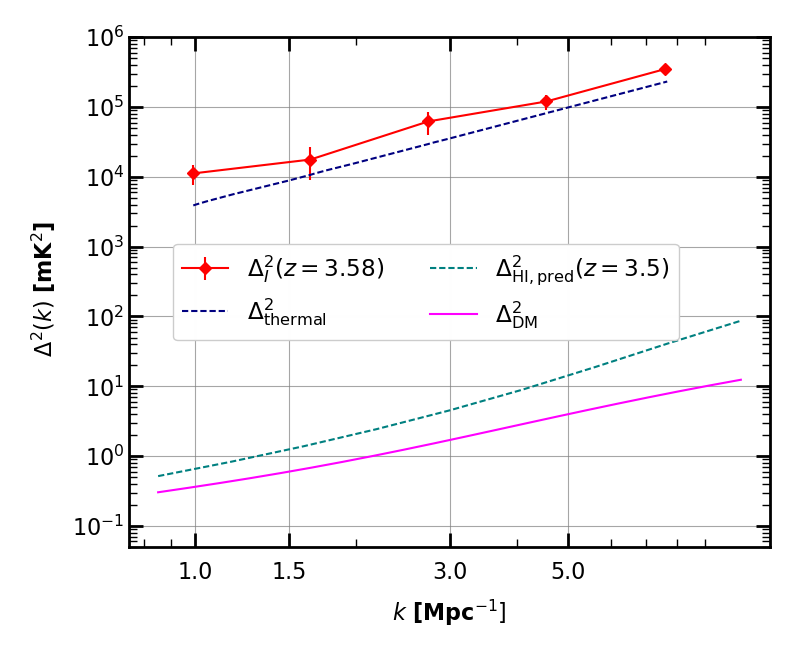} 
       
%\end{tabular}      
         
    \caption{The spherically averaged dimensionless power spectrum, $\Delta^{2}_{I}$,  at $z=3.58$ in red. The dashed line in navy is the theoretical estimate of the thermal noise power. The bottom dashed line in teal is the theoretical prediction of H{\sc i} power spectrum at $z=3.5$ taken from \citet{Debanjan2016MNRAS.460.4310S}. The last magenta line is the dark matter (DM) power spectrum at $z=3.58$ estimated using CAMB.}
    \label{3d_PS_310}
\end{figure}

We choose four 8 MHz sub-bands from the the entire observed  bandwidth which are relatively less contaminated by RFI and estimate the cylindrically and spherically averaged power spectrum.  In Fig. \ref{PS_310}, we show the cylindrically averaged power spectrum estimated from a 8 MHz sub-band around 310 MHz corresponding to redshift $z=3.58$. Note that, the first and last 4 MHz of the 200 MHz bandwidth have been flagged. Hence, this is the power spectrum of the highest redshift ($z=3.58$) probed in this analysis. We find that the spectrally smooth foregrounds coupled with chromatic instrument response are confined in a `wedge' shape region in this 2D space.  This cylindrical averaged power spectrum is useful to identify the $[k_{\perp}, k_{\parallel}]$-modes devoid of foreground contamination. The black line in Fig. \ref{PS_310} shows the upper bound of foreground contaminated modes within the horizon ($\tau = |\bold U|\sin{90^{\circ}}/c$) limit. 

\begin{figure}[htp]
    \centering
   
%\begin{tabular}{cc}
        
       \includegraphics[width=\columnwidth,height=2.8in]{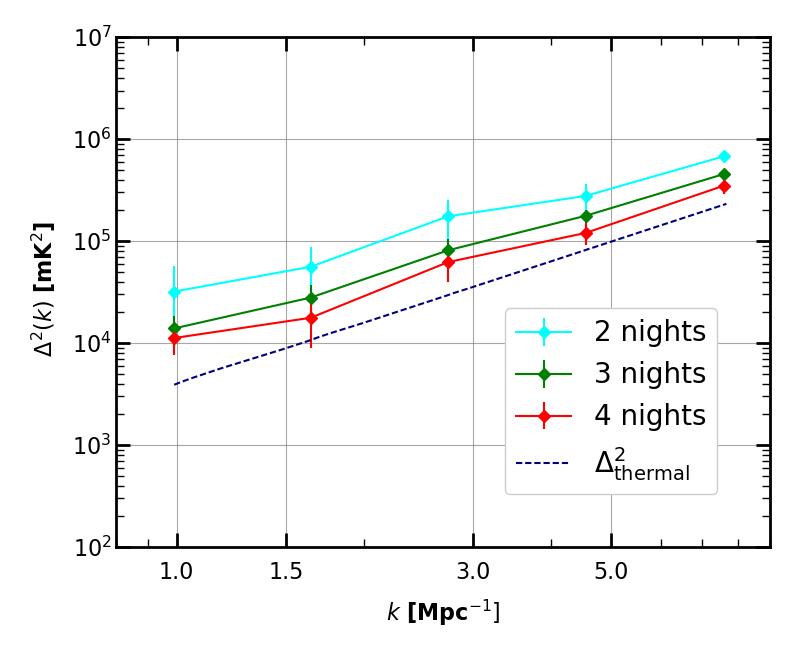} 
       
%\end{tabular}      
         
    \caption{The spherically averaged  power spectrum at $z=3.58$, after combining different night's data coherently. The bottom navy dashed line is the theoretical thermal noise power spectrum for the whole 4 nights data. }
    \label{3d_PS_compare}
\end{figure}

Above the horizon line we find a  region in $[k_{\perp}, k_{\parallel}]$-space where the power is 2-3 orders of magnitude less than the power inside the horizon limit.  We choose the modes bounded by the white dashed curve (see Fig. \ref{PS_310}) by visually inspecting the less foreground contaminated region and estimate the spherically averaged power spectrum using those modes. The dimensionless  spherically averaged stokes-$I$ power spectrum ($\Delta^{2}_{I}$) in $\mathrm{mK^{2}}$ along with $2\sigma$ error bars at $z=3.58$ are being shown in Fig. \ref{3d_PS_310}.   We also estimate the theoretical thermal noise power ($\Delta^{2}_{\mathrm{thermal}}$), including the flagging,  using the uGMRT baseline distribution, bandwidth, integration time and system temperature ($T_{sys}$) \citep{McQuinn2006ApJ...653..815M,Parsons2012BApJ...753...81P}. The quantity $G/T_{sys}$ as a function of frequency for uGMRT is given as a polynomial \footnote{\url{http://www.ncra.tifr.res.in:8081/~secr-ops/etc/etc_help.pdf}}, where $G$ is the antenna gain \citep{Yashwant2017CSci..113..707G}. We estimate the $T_{sys}$ at the central frequency of the sub-band using the polynomial and also correct for the sky-temperature at the corresponding frequency.  The bottom green dash line shows the theoretical prediction of H{\sc i} 21 cm power spectrum at redshift $z=3.5$ taken from \citep{Debanjan2016MNRAS.460.4310S}. We find that estimated $\Delta^{2}_{I}$  is close to  $\Delta^{2}_{\mathrm{thermal}}$ at the $k$ modes probed here.  However, the measured $\Delta^{2}_{I}$ is  nearly four orders of magnitude higher than the theoretical expectation of H{\sc i} 21 cm power spectrum close to redshift $z=3.58$. Hence, this analysis with $\sim$ 13 hours of on source data (before flagging) put an  upper limit on  the post-EoR H{\sc i} power spectrum and limited by thermal noise.

 The data was observed over 4 nights and coherently added in the $uv$-domain within $uv$-cells. We show in Fig. \ref{3d_PS_compare} the spherically averaged power spectrum after combining different night's data set successively. We find a systematic decrease in power upon combining more data in $uv$-domain. This proves that the coherent addition of data in $uv$-domain and the methodology of our power spectrum estimation are consistent with theoretical expectation. 

\begin{figure}[htp]
    \centering
   
%\begin{tabular}{cc}
        
       \includegraphics[width=\columnwidth,height=2.8in]{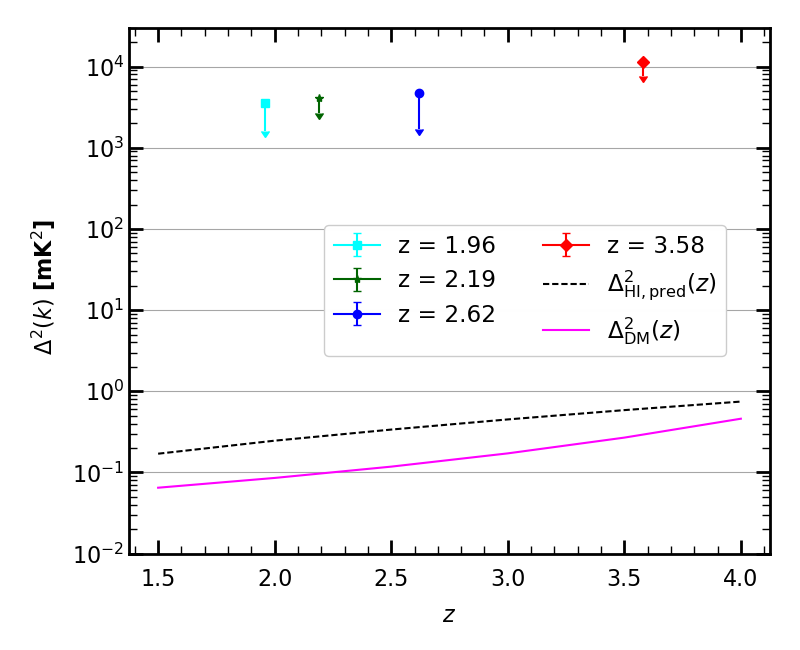} 
       
%\end{tabular}      
         
    \caption{The lowest limit on spherically averaged power spectrum for different redshifts at $k=1.0$. The black dashed curve shows the theoretical prediction of H{\sc i} power spectrum as a function of $z$ at $k=1.0$ $\mathrm{Mpc}^{-1}$ \citep{Debanjan2016MNRAS.460.4310S}. The bottom magenta curve is the dark matter power spectrum as a function of redshift.}
    \label{ps_vs_z}
\end{figure}

We also choose three other less RFI contaminated 8 MHz sub-bands around 392 MHz ($z=2.62$), 444 MHz ($z=2.19$) and 479 MHz ($z=1.96$) and estimate the power spectrum following the same procedure discussed above. The cylindrically and spherically averaged  power spectrum plots are being shown in Appendix \ref{app:multi_redshift}. The lowest limits, at $k \sim 1.0$ $\mathrm{Mpc}^{-1}$, on  spherically averaged 21 cm power spectrum are (58.87 $\mathrm{mK})^{2}$, (61.49 $\mathrm{mK})^{2}$, (60.89 $\mathrm{mK})^{2}$, (105.85 $\mathrm{mK})^{2}$  at $z = 1.96,2.19,2.62$ and $3.58$, respectively  and shown  in Fig. \ref{ps_vs_z}.  The values of $\Delta^{2}_{I}$ along with the $2\sigma$ error bars for all $k$ at different redshifts are mentioned in Table \ref{values} (Appendix \ref{tabulated}).  

 Note that, there are several different ways which  may contaminate the foreground free cosmological window above the ‘wedge’, like residual calibration errors, polarization leakage, ionospheric effects, the variation of beam, etc and affect the estimation of the \hi\ 21 cm power spectrum  \citep{Gehlot2018MNRAS.478.1484G,Joseph2020MNRAS.492.2017J,Jais2020MNRAS.495.3683K}. Although the estimated power spectrum for different redshift bins are close to the thermal noise,  the resultant power spectrum may still be affected by any plausible residual systematics. We will analyze any plausible contamination of the signal window due to different systematics in detail and present in future work.
 
\section{Constraints on \texorpdfstring{[$\Omega_{\mathrm{H{\sc i}}} \lowercase{b}_{\mathrm{H{\sc i}}}$]}{}}

The main observable in 21 cm intensity mapping experiments is the power spectrum of 21 cm brightness temperature fluctuation and given by the expression \citep{Battye2012arXiv1209.1041B,Anderson2018MNRAS.476.3382A}:

\begin{equation}
 %   \begin{math}
 \label{eqn5}
    \Delta^{2}_{\mathrm{H{\sc I}}} (k,z) = \overline{T}(z)^{2} [b_{\mathrm{H{\sc I}}} (k,z)]^{2} \frac{k^{3}P_{\mathrm{DM}}(k,z)}{2 \pi^{2}},
%\end{math}
\end{equation}

where the mean brightness temperature $\overline{T}(z)^{2}$ is given by \citep{Anderson2018MNRAS.476.3382A},

\begin{equation}
\label{eqn6}
\begin{split}
    \overline{T}(z) \simeq 0.39 \frac{\Omega_{\mathrm{H{\sc I}}}(z)}{10^{-3}} \Big[\frac{\Omega_{\mathrm{m}}+\Omega_{\Lambda}(1+z)^{-3}}{0.29}\Big]^{-1/2} \\  \Big[\frac{(1+z)}{2.5}\Big]^{1/2} \mathrm{mK},
\end{split}    
\end{equation}

where $b_{\mathrm{H{\sc I}}} (k,z)$ is the  \hi\  bias and $P_{\mathrm{DM}}(k,z)$ is the dark matter power spectrum. 

We use CAMB \footnote{\url{https://camb.info/}} code to generate the dark matter power spectrum  at any given redshift for the $k$ range probed here. 
 After combining Eqns. \ref{eqn5} and \ref{eqn6} and using  the best limit on H{\sc i} power spectrum at $k=1.0$ $\mathrm{Mpc}^{-1}$, the estimated upper limit  on [$\Omega_{\mathrm{H{\sc I}}} b_{\mathrm{H{\sc I}}}$] are 0.09,0.11,0.12,0.24 at $z=1.96,2.19,2.62,3.58$, respectively. Theoretical prediction shows the value of  [$b_{\mathrm{H{\sc I}}}$] $\sim 1-1.5$ at $k\sim 1$  Mpc$^{-1}$ for the redshift range probed here \citep{Debanjan2016MNRAS.460.4310S}. Previous observation of DLAs suggests that [$\Omega_{\mathrm{H{\sc I}}}$] $\sim 5\times 10^{-4}$ at this redshifts \citep{Hamsa2015MNRAS.447.3745P}, which is still 150 times smaller than our present limit.  

\section{Summary}

In this analysis, using 13 hours uGMRT observation of the  ELAIS N1 field, we put limits on post-EoR H{\sc i} power spectrum at redshifts $z=1.96,2.19,2.62,3.58$. This is the first attempt to estimate the statistical feature of post-EoR H{\sc i} signal via auto power spectrum using interferomteric data at this redshift range using uGMRT.

 To maintain signal ergodicity, we divide the whole 200 MHz band into 8 MHz sub-bands and choose 4 such less RFI contaminated sub-bands  to put limits on H{\sc i} power spectrum. We coherently added the data in $uv$-domain and estimate the  cosmological \hi\ 21 cm power spectrum  with properly accounting for the positive noise bias. We use 1D $\it{`CLEAN'}$ algorithm  to mitigate the foreground spillover beyond horizon limit caused by missing channels due to RFI flagging. We estimate the cylindrically averaged  power spectrum  using proper scaling factors and find that spectrally smooth foregrounds coupled with chromatic instrument response is contained within a `wedge' shape region inside the horizon limit. Using the  modes less contaminated by the foregrounds above the horizon limit, we estimate the spherically averaged power spectrum.  The upper limits at $k=1.0$ $\mathrm{Mpc}^{-1}$ on  spherically averaged 21 cm power spectrum are (58.87 $\mathrm{mK})^{2}$, (61.49 $\mathrm{mK})^{2}$, (60.89 $\mathrm{mK})^{2}$, (105.85 $\mathrm{mK})^{2}$ and the corresponding  limit on the quantity  [$\Omega_{\mathrm{H{\sc I}}} b_{\mathrm{H{\sc I}}}$]  are 0.09,0.11,0.12,0.24 at $z=1.96,2.19,2.62,3.58$, respectively.

  This analysis with the uGMRT observation is a first attempt to characterize the fluctuations in the post-EoR H{\sc i} signal at these high redshifts using the `foreground avoidance' technique and demonstrates that the post-EoR \hi\ 21 cm power spectrum can be detected with more than 5000 hours of observation using uGMRT. We also constrain the product of the astrophysical quantities $b_{\mathrm{H{\sc I}}}$ and $\Omega_{\mathrm{H{\sc I}}}$, which will be helpful to characterize the uncertainties associated with the measurement of \hi\ power spectrum  for ongoing and future \hi\ 21 cm intensity mapping experiments like, OWFA \footnote{\url{http://www.ncra.tifr.res.in/ncra/ort}}, CHIME \footnote{\url{https://chime-experiment.ca/en}} , TIANLAI \footnote{\url{https://tianlai.bao.ac.cn}}, HIRAX \footnote{\url{https://hirax.ukzn.ac.za/}} and the SKA-1 mid \footnote{\url{https://www.skatelescope.org/mfaa/}}. \\

{\bf ACKNOWLEDGEMENTS}

We  thank the anonymous referee  for helpful comments and suggestions that have helped to improve this work. We  thank the staff of GMRT for making this observation possible. GMRT is run by National Centre for Radio Astrophysics of the Tata Institute of Fundamental Research. AC would like to thank DST for INSPIRE fellowship. AD would like to acknowledge the support of EMR-II under CSIR No. 03(1461)/19.

\bibliography{upper_limit}
\bibliographystyle{aasjournal}

\appendix

\counterwithin{figure}{section}
\counterwithin{table}{section}

\section{2D and 3D power spectrum at \texorpdfstring{$z = 1.96,2.19,2.62$}{}} \label{app:multi_redshift}

\begin{figure*}[htp]
\centering
   
\begin{tabular}{c}
        
       \includegraphics[width=6.5in]{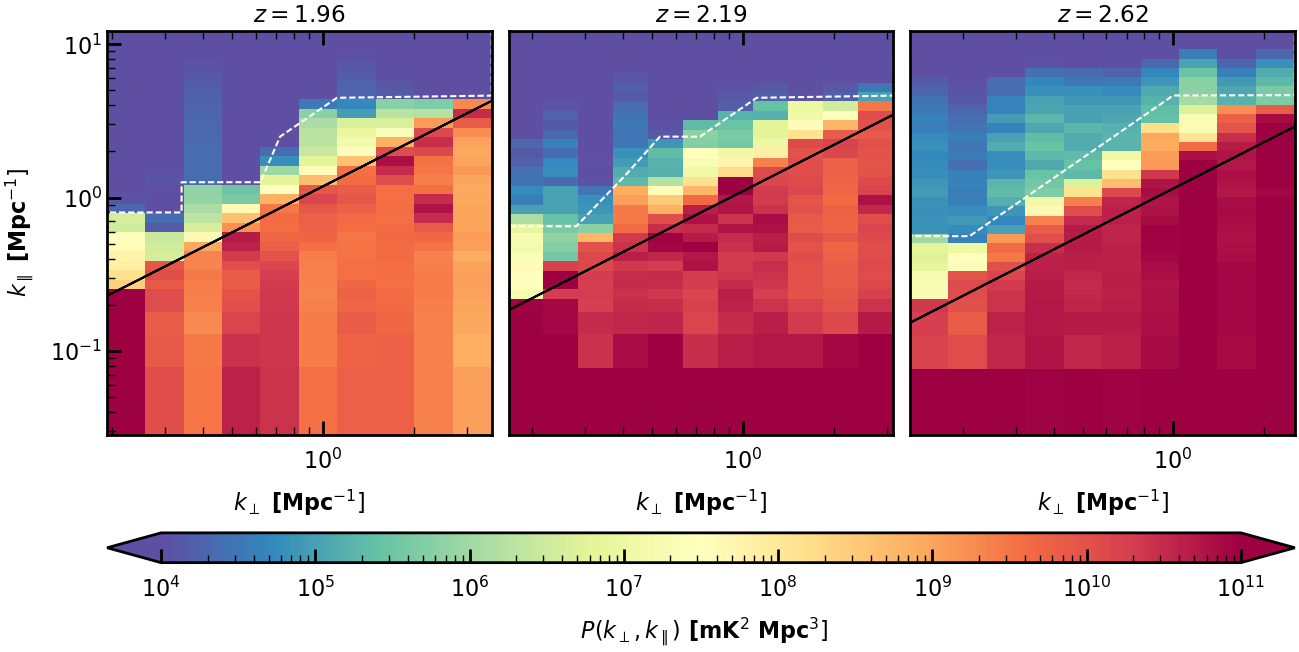} \\
       \includegraphics[width=6.5in]{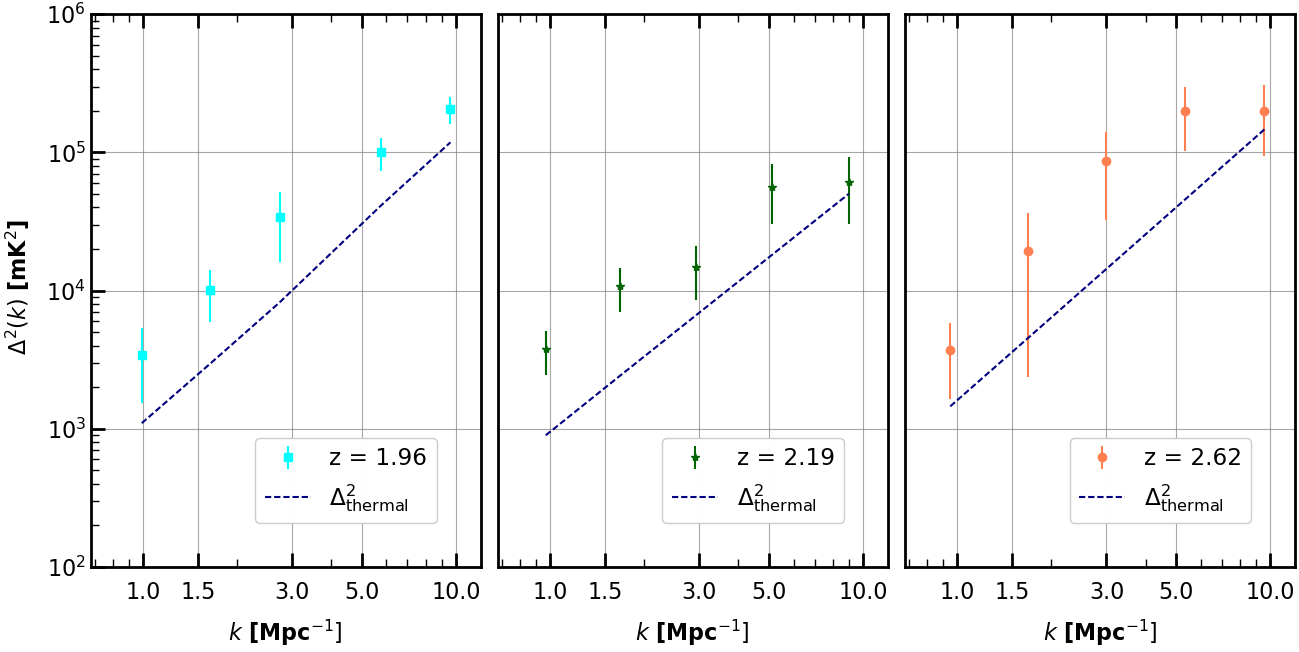}

\end{tabular}      
         
    \caption{ The cylindrically  averaged 2D power spectrum (upper panel) and spherically averaged 3D power spectrum (lower panel) for redshifts $z$ = 1.96 (left column), 2.19 (middle column) and 2.62 (right column).}
    \label{app:wedge_all}
\end{figure*} 

\section{Tabulated power spectrum values at different redshifts}
\label{tabulated}

\begin{table*}[htp]
%\centering
\caption{Upper limit values estimated power spectrum for redshifts \texorpdfstring{$z = 1.96, 2.19, 2.62, 3.58$}{}}
\label{values}

\begin{floatrow}[2]

\begin{tabular}{|c|c|c|}
\hline
\multicolumn{3}{|c|}{$\mathbf{z = 1.96}$}\\

\hline
 $k$ &   $\Delta^{2}_{I}$ &  $\Delta^{2}_{I,err}$ \\
 $\mathrm{Mpc}^{-1}$ & mK$^{2}$ & mK$^{2}$ \\
\hline

 0.99 &  $\mathbf{(58.57)^{2}}$ & $(43.54)^{2}$ \\
 1.64 & $(100.08)^{2}$ & $(64.11)^{2}$ \\
 2.73 & $(184.25)^{2}$ & $(133.92)^{2}$ \\
 5.74 & $(316.94)^{2}$ & $(162.87)^{2}$ \\
 9.60 & $(452.71)^{2}$ & $(212.27)^{2}$ \\
     
\hline     
\end{tabular}

\quad

\begin{tabular}{|c|c|c|}
\hline
\multicolumn{3}{|c|}{$\mathbf{z = 2.19}$} \\

\hline
 $k$ &   $\Delta^{2}_{I}$ &  $\Delta^{2}_{I,err}$ \\
 $\mathrm{Mpc}^{-1}$ & mK$^{2}$ & mK$^{2}$ \\
\hline

 0.97 &  $\mathbf{(61.49)^{2}}$ & $(36.50)^{2}$ \\
 1.67 & $(103.84)^{2}$ & $(61.38)^{2}$ \\
 2.92 & $(121.89)^{2}$ & $(79.52)^{2}$ \\
 5.12 & $(236.55)^{2}$ & $(160.70)^{2}$ \\
 9.02 & $(247.93)^{2}$ & $(176.01)^{2}$ \\
     
\hline     
\end{tabular}

\end{floatrow}

\vspace{1cm}

\begin{floatrow}[2]

\begin{tabular}{|c|c|c|}
\hline
\multicolumn{3}{|c|}{$\mathbf{z = 2.62}$} \\

\hline
 $k$ &   $\Delta^{2}_{I}$ &  $\Delta^{2}_{I,err}$ \\
 $\mathrm{Mpc}^{-1}$ & mK$^{2}$ & mK$^{2}$ \\
\hline

 0.95 &  $\mathbf{(60.89)^{2}}$ & $(45.49)^{2}$ \\
 1.68 & $(139.47)^{2}$ & $(130.72)^{2}$ \\
 2.99 & $(294.13)^{2}$ & $(232.13)^{2}$ \\
 5.36 & $(446.21)^{2}$ & $(311.65)^{2}$ \\
 9.59 & $(447.51)^{2}$ & $(325.50)^{2}$ \\
     
\hline     
\end{tabular}

\quad

\begin{tabular}{|c|c|c|}
\hline
\multicolumn{3}{|c|}{$\mathbf{z = 3.58}$} \\

\hline
 $k$ &   $\Delta^{2}_{I}$ &  $\Delta^{2}_{I,err}$ \\
 $\mathrm{Mpc}^{-1}$ & mK$^{2}$ & mK$^{2}$ \\
\hline

 0.99 &  $\mathbf{(105.85)^{2}}$ & $(60.34)^{2}$ \\
 1.64 & $(133.06)^{2}$ & $(93.46)^{2}$ \\
 2.73 & $(249.49)^{2}$ & $(150.38)^{2}$ \\
 4.54 & $(346.92)^{2}$ & $(170.44)^{2}$ \\
 7.58 & $(590.82)^{2}$ & $(241.76)^{2}$ \\
     
\hline     
\end{tabular}
\end{floatrow}

\end{table*}

%\listofchanges
\end{document}